\documentclass[12pt]{article}
\usepackage{amssymb,amsmath,amsthm,latexsym}

\newtheorem{thm}{Theorem}[section]

\newtheorem{lem}[thm]{Lemma}
\newtheorem{cor}[thm]{Corollary}
\newtheorem{exam}{Example}
\newtheorem{defi}[thm]{Definition}

\newcommand{\pf}{{\bf Proof. \ }}

\begin{document}

\title{Binary LCD Codes from $\mathbb{Z}_2\mathbb{Z}_2[u]$}
\author{Peng Hu\\
 School of Mathematics and Physics, \\
 Hubei Polytechnic University  \\
 Huangshi, Hubei 435003, China, \\
 {Email: \tt HPhblg@126.com} \\
Xiusheng Liu\\
 School of Mathematics and Physics, \\
 Hubei Polytechnic University  \\
 Huangshi, Hubei 435003, China, \\
{Email: \tt lxs6682@163.com} \\}
\maketitle


\begin{abstract} Linear complementary dual (LCD) codes over finite fields are linear codes satisfying $C\cap C^{\perp}=\{0\}$. We generalize the LCD codes over finite fields to $\mathbb{Z}_2\mathbb{Z}_2[u]$-LCD codes over the ring $\mathbf{Z}_2\times(\mathbf{Z}_2+u\mathbf{Z}_2)$.  Under suitable conditions, $\mathbb{Z}_2\mathbb{Z}_2[u]$-linear codes that are $\mathbb{Z}_2\mathbb{Z}_2[u]$-LCD codes are characterized.  We then prove that the binary image of a $\mathbb{Z}_2\mathbb{Z}_2[u]$-LCD code is a binary LCD code. Finally,  by means of these conditions,  we construct new binary LCD codes using $\mathbb{Z}_2\mathbb{Z}_2[u]$-LCD codes, most of which have better
parameters than current binary LCD codes available.
\end{abstract}


\bf Key Words\rm: $\mathbb{Z}_2\mathbb{Z}_2[u]$-linear codes; $\mathbb{Z}_2\mathbb{Z}_2[u]$-LCD codes;   self-orthogonal codes

\bf Mathematics Subject Classification (2010) \rm 94B05 11T71

\section{Introduction}
Linear complementary dual codes (which is abbreviated to LCD codes) are linear codes that meet their dual trivially. These codes were introduced by Massey in \cite{Massey} and showed that asymptotically good LCD codes exist, and provide an optimum linear coding solution for the two-user binary adder channel. They are also used in counter measure to passive and active side channel analyses on embedded cryto-systems(See\cite{Carlet}). In recently, Guenda, Jitman and Gulliver investigated an application of LCD codes in constructed good entanglement-assisted quantum error correcting codes \cite{Guend}.

Yang and Massy in \cite{Yang} showed that a necessary and sufficient condition for a cyclic code of length $n$ over finite fields to be an LCD code is that the generator polynomial $g(x)$ is self-reciprocal and all the monic irreducible factors of $g(x)$ have the same multiplicity in $g(x)$ as in $x^n-1$. In \cite{Sendrier}, Sendrier indicated that linear codes with complementary-duals  meet the asymptotic Gilbert-Varshamov bound. Dougbherty, Kim, Ozkaya, Sok and Sol$\acute{e}$ developed a linear programming bound on the largest size of an LCD code of given length and minimum distance \cite{Dougherty}.  Ding, C. Li and S. Li in \cite{Ding} constructed  LCD BCH codes.  Liu, Fan and Liu \cite{Liu2} introduced the so-called $k$-Galois LCD codes which include the usual LCD codes and Hermitian LCD codes as two special cases, and gave sufficient and necessary conditions for a code to be a $k$-Galois LCD code. In recently, Carlet et al. solved the problem
of the existence of $q$-ary LCD MDS codes for Euclidean case \cite{Carlet1}, they also introduced a general construction of LCD codes from any linear codes. Further more, they showed that
any linear code over $\mathbb{F}_q$  ($q > 3$) is equivalent to an Euclidean LCD code and any linear code
over $\mathbb{F}_{q^2}$  ($q > 2$) is equivalent to a Hermitian LCD code \cite{Carlet2}.  But, in \cite{Pang}, B. Pang, S. Zhu, and X. Kai show that LCD codes are not equivalent to linear codes over $\mathbb{F}_2$. This motivates us to study binary LCD codes.

We finish this introduction with a description of each section in this paper. Section 2 reviews the basics about $\mathbb{Z}_2\mathbb{Z}_2[u]$-linear codes and LCD codes. In Section 3,  we propose new constructions of binary LCD codes by using $\mathbb{Z}_2\mathbb{Z}_2[u]$-LCD codes,  and, in Section 4, using methods of the section 3, concrete examples are presented to construct good parameters binary LCD codes.  Finally, a brief summary of this work is described in Section 5.

\section{Preliminaries}
In order for the exposition in this paper to be self-contained, we introduce some basic concepts and results about $\mathbb{Z}_2\mathbb{Z}_2[u]$-linear codes and LCD codes. For more details, we refer to  \cite{Aydo},\cite{Massey},\cite{Huffman}.

Starting from this section till the end of this paper, we denote the ring $\mathbb{Z}_2+u\mathbb{Z}_2$ by $R$, where $u^2=0$. It is easy to see that the ring $\mathbb{Z}_2$ is a subring of the ring $R$.  We  define the set
$$\mathbb{Z}_2\times R=\{(a,b)\mid a \in\mathbb{Z}_2~\mathrm{and}~b\in R\}.$$

Let $r=s+ut\in R$. Define the map $\theta:  R \longrightarrow \mathbb{Z}_2$, such that $\theta(r)=s$.
So, $\theta(0)=0,\theta(1)=1,\theta(u)=0$, and $\theta(1+u)=1$. Obviously, the mapping $\theta$ is a homomorphism from ring $R$ to $\mathbb{Z}_2$. Now, for any element $r\in R$, define an $R$-scalar multiplication on  $\mathbb{Z}_2\times R$ as
$$r(a,b)=(\theta(r)a,rb).$$

Furthermore, this multiplication can be extended naturally to $\mathbb{Z}_2^{\alpha}\times R^{\beta}$ as follows. For any $r\in R$ and $\mathbf{v}=(a_1,\ldots,a_{\alpha},b_1,\ldots,b_{\beta})\in \mathbb{Z}_2^{\alpha}\times R^{\beta}$ define
$$r(a_1,\ldots,a_{\alpha},b_1,\ldots,b_{\beta})=(\theta(r)a_1,\ldots,\theta(r)a_{\alpha},rb_1,\ldots,rb_{\beta}). $$

\begin{defi} A $\mathbb{Z}_2\mathbb{Z}_2[u]$-linear code $C$ is a non-empty $R$-submodule of $\mathbb{Z}_2^{\alpha}\times R^{\beta}$.  If $C \subset\mathbb{Z}_2^{\alpha}\times R^{\beta}$ is a $\mathbb{Z}_2\mathbb{Z}_2[u]$-linear code, group isomorphic to $\mathbb{Z}_2^{k_0+k_1}\times \mathbb{Z}_2^{k_2}$, then C is called a $\mathbb{Z}_2\mathbb{Z}_2[u]$-linear code of type $(\alpha,\beta,k_0,k_1,k_2)$ where $k_0,k_1$ and $k_2$ are defined above.
\end{defi}

Now we recall the definition of the Gray map on $\mathbb{Z}_2^{\alpha}\times R^{\beta}$. Observe that any element $r \in R$ can be expressed as $r = s +ut$, where
$r , q \in\mathbb{Z}_2$ . The Gray map on $\mathbb{Z}_2^{\alpha}\times R^{\beta}$, defined in \cite{Aydo}, can be written as
$$\Phi:\quad \mathbb{Z}_2^{\alpha}\times R^{\beta}  \longrightarrow \mathbb{Z}_2^n,$$
$$\Phi(a_1,\ldots,a_{\alpha},s_1 +ut_1,\ldots,s_{\beta} +ut_{\beta})=(a_1,\ldots,a_{\alpha},t_1,\ldots,t_{\beta},s_1 +t_1,\ldots,s_{\beta} +t_{\beta}).$$

Note that the binary image $\Phi(C)$ of a $\mathbb{Z}_2\mathbb{Z}_2[u]$-linear code $C$ of type $(\alpha,\beta,k_0,k_1,k_2)$ is a binary linear code of length $n=\alpha+2\beta$ and size $2^{k_0+2k_1+k_2}$.  It is also called a $\mathbb{Z}_2\mathbb{Z}_2[u]$-linear code.

The Lee weight $W_L(r)$ of the element $r\in R$ is defined as
\[
W_L(r)=\left\{
           \begin{array}{ll}
             0, & \hbox{if\quad $r=0$;} \\
             1, & \hbox{if\quad $r=1,1+u$;} \\
             2, & \hbox{if\quad $r=u$.}
           \end{array}
         \right.
\]
Denote by $W_L(\mathbf{b})$ the Lee weight of $\mathbf{b}\in R^{\beta}$, which is the rational sum of Lee weights of the coordinates of $\mathbf{b}$.  For a vector $\mathbf{v}=(\mathbf{a}|\mathbf{b})\in\mathbb{Z}_2^{\alpha}\times R^{\beta}$, the Lee weight  of $\mathbf{v}$ is defined as $W_L(\mathbf{v})=W_H(\mathbf{a})+W_L(\mathbf{b})$, where $W_H(\mathbf{a})$ is the Hamming weight of $\mathbf{a}\in\mathbb{Z}_2^{\alpha}$.
The minimum Lee distance $d_L(C)$ of a $\mathbb{Z}_2\mathbb{Z}_2[u]$-linear code $C$ is the smallest nonzero Lee distance between all pairs
of distinct codewords of $C$. Obviously, $d_L(C)=W_H(\Phi(C))$.

The inner product in $\mathbb{Z}_2^{\alpha}\times R^{\beta}$, defined in \cite{Aydo}, can be written as
$$[ \mathbf{v},\mathbf{w}]=u\sum_{i=1}^{\alpha}a_{i}b_{i}+\sum_{j=1}^{\beta}x_{j}y_{j}\in R.$$
where $\mathbf{v}=(a_1,a_2,\ldots,a_{\alpha},x_1,\ldots,x_{\beta})$ and $\mathbf{w}=(b_1,b_2,\ldots,b_{\alpha},y_1,\ldots,y_{\beta})$ .

The dual code of a $\mathbb{Z}_2\mathbb{Z}_2[u]$-linear code $C$ is defined in the standard way by
$$C^{\perp}=\{\mathbf{v}\in\mathbb{Z}_2^{\alpha}\times R^{\beta}\mid [ \mathbf{v},\mathbf{w}]=0~\mathrm{for}~\mathrm{all}~\mathbf{w}\in C\}.$$
We say that the code $C$ is self-orthogonal, if $C\subset C ^{\perp}$ and self-dual, if $C=C^{\perp}$.

A $\mathbb{Z}_2\mathbb{Z}_2[u]$-linear code $C$ is called $\mathbb{Z}_2\mathbb{Z}_2[u]$-LCD  if $C^{\perp}\cap C=\{\mathbf{0}\} $.

\begin{lem} \label{le:3.1} If $C$ is a $\mathbb{Z}_2\mathbb{Z}_2[u]$-linear code of type $(\alpha,\beta,k_0,k_1,k_2)$, then  $|C|\cdot|C^{\perp}|=2^{\alpha+2\beta}$.
\end{lem}
\pf  According to [2, Theorem 2.2 and Theorem 2.7], $C$ is permutation equivalent to a $\mathbb{Z}_2\mathbb{Z}_2[u]$-linear code with the standard form matrix
$$G=\left(
\begin{array}{ccccc}
I_{k_0}& A_{1}&0& 0 &uP\\
0 &S &I_{k_1} &A &B_1+uB_2\\
0 &0 &0 &uI_{k_2}& uD\\
\end{array}
\right),$$
and, its dual code $C^{\perp}$ with the generator matrix
$$H=\left(
\begin{array}{ccccc}
A_{1}^{T}&I_{\alpha-k_0} &-uS^T& 0 &0\\
-P^{T} &0&-(B_1+uB_2)^T+D^TA^T & -D^T&I_{\beta-k_1-k_2}\\
0 &0 &-uA^T &uI_{k_2}&0\\
\end{array}
\right).$$
Thus, $|C|\cdot|C^{\perp}|=2^{k_0+2k_1+k_2}2^{\alpha-k_0+2(\beta-k_1-k_2)+k_2}=2^{\alpha+2\beta}$.

\begin{thm}
Let $C$ be a $\mathbb{Z}_2\mathbb{Z}_2[u]$-linear code of type $(\alpha,\beta,k_0,k_1,k_2)$. Then
$\Phi(C^\bot)=\Phi(C)^\bot$,  where $\Phi(C)^\bot$ is
the ordinary dual of $\Phi(C)$ as a binary code.
\end{thm}
\pf Let $\mathbf{x}=(\mathbf{a},\mathbf{b}+u\mathbf{c}), \mathbf{y}=(\mathbf{d}, \mathbf{e}+u\mathbf{f})\in C$ be two codewords, where $\mathbf{a,
d}\in\mathbb{ Z}_2^{\alpha}$, and $\mathbf{b,
c,e,f}\in\mathbb{ Z}_2^{\beta}$. Then
$$[\mathbf{x, y}]= [\mathbf{b,e}]+u([\mathbf{a, d}]+ [\mathbf{b, f}]+[\mathbf{c,e}]),$$
and
$$[\Phi(\mathbf{x}), \Phi(\mathbf{y})]= \mathbf{[a, d]+ [c,f]+[b,e]+[b,f]+[c,e]+[c,f]=[b,e]+([a, d]+ [b, f]+[c,e]}).$$

It is easy to check that $[\mathbf{x, y}]=0$ implies $[\Phi(\mathbf{x}), \Phi(\mathbf{y})]=0.$ Therefore, $$\Phi(C^\bot)\subset\Phi(C)^\bot.\eqno(2.1)$$

But, by the definition of $\Phi$, $\Phi(C)$ is a binary linear code of length $\alpha+2\beta$ of size $|C|$.
 So, by the usual properties of the dual of binary codes, we know that $|\Phi(C)^\bot|=\frac{2^{\alpha+2\beta}}{|C|}.$

On the other hand, by Lemma \ref{le:3.1}, we have $|\Phi(C^\bot)|=|C^\bot|=\frac{2^{\alpha+2\beta}}{|C|}$.

Thus
 $$|\Phi(C^\bot)|=|\Phi(C)^\bot|\eqno(2.2)$$
 Combining (2.1) and (2.2), we get the desired equality.
\qed

\section{ $\mathbb{Z}_2\mathbb{Z}_2[u]$-LCD Codes}

It  is easy to see that $R$ is a local Frobenius ring with unique maximal ideal $\mathfrak{m}=\{0,u\}$.

We begin with some definitions and lemmas with respect to vectors in  $\mathbb{Z}_2^{\alpha}\times R^{\beta}$.

\begin{defi} \label {de:3.1}Let $\mathbf{w}_1,\ldots,\mathbf{w}_k$ be non-zero vectors in $\mathbb{Z}_2^{\alpha}\times R^{\beta}$ . Then $\mathbf{w}_1,\ldots,\mathbf{w}_k$ are $R$-independent if $\sum_{j=1}^k\delta_j\mathbf{w}_j=0$
implies that $\delta_j\mathbf{w}_j=0 $ for all $j$, where $\delta_j\in R$.
\end{defi}

Following Definition \ref{de:3.1}, we can easily get the following lemma.
\begin{lem} \label{le:3.2} If the non-zero vectors $\mathbf{w}_1 ,\ldots,\mathbf{w}_s$ in $\mathbb{Z}^{\alpha}\times R^{\beta}$
are $R$-independent and  $\sum_{j=1}^s\delta_j\mathbf{w}_j=0$, then  $\delta_j \in\mathfrak{ m }$ for all $j$.
\end{lem}

\pf Since $\sum_{j=1}^s\delta_j\mathbf{w}_j=0$. Then $\delta_j \mathbf{w}_j = 0 $ for all $j$. If $\delta_j \notin\mathfrak{m}$ for some $j$, then $\delta_j$ is a unit, and this implies that $\mathbf{w}_j = 0$, which is a contradiction.
\qed

Let $\mathbf{w}_1,\ldots,\mathbf{w}_s$  be vectors in $\mathbb{Z}^{\alpha}\times R^{\beta}$.
As usual, we denote the set of all linear combinations
of $\mathbf{w}_1,\ldots,\mathbf{w}_s$ by $\langle \mathbf{w}_1,\ldots,\mathbf{w}_s \rangle$.

\begin{lem} If the non-zero vectors $\mathbf{w}_1 ,\ldots,\mathbf{w}_s$ in $\mathbb{Z}^{\alpha}\times R^{\beta}$
are $R$-independent, then  $\mathbf{w}_1 ,\ldots,\mathbf{w}_s$ none is  a linear combination of the other
vectors.
\end{lem}
\pf Without loss of generality, suppose $\mathbf{w}_s$ can be written as a linear combination of the other
vectors. Then
$$\mathbf{w}_s =\sum_{j=1}^{s-1}\alpha_j\mathbf{w}_j.$$
This gives
$$\alpha_1 \mathbf{w}_1 + \cdots + \alpha _{s-1}\mathbf{ w}_{s-1}+(-1)\mathbf{w}_s = 0,$$
which is a contradiction to Lemma \ref{le:3.2}.

Central to the study of algebraic coding theory is the concept of a code generator matrix. The rows of the generator matrix form a basis of the code. We shall now give a definition of a basis of a $\mathbb{Z}_2\mathbb{Z}_2[u]$-linear code.

\begin{defi} \label {de:4} Let $C\neq\{\mathbf{0}\}$ be a $\mathbb{Z}_2\mathbb{Z}_2[u]$-linear code. The non-zero codewords $\mathbf{c}_1, \mathbf{c}_2,\ldots, \mathbf{c}_k$ are called a basis of $C$ if they are $R$-independent and generate $C$. Set $G=\begin{pmatrix}\mathbf{c}_1\\\mathbf{c}_2\\\vdots\\\mathbf{c}_k\end{pmatrix}$. We say that $G$ is a generator matrix of $C$.
\end{defi}

Given a $k\times n$ matrix $G$, we denote by $G(i,:)$ the  $i$-th row of $G$.

In terms of the generator matrix,
we now  give a sufficient condition for   a  $\mathbb{Z}_2\mathbb{Z}_2[u]$-linear code  to be LCD.
\begin{thm} \label{th:A}
Let  $C\neq\{\mathbf{0}\}$ be a  $\mathbb{Z}_2\mathbb{Z}_2[u]$-linear code with generator matrix  $G$. If the $k\times k$ matrix $GG^{T}$ is invertible, then $C$ is a $\mathbb{Z}_2\mathbb{Z}_2[u]$-LCD code, where $k$ is the number of rows of $G$.
\end{thm}
\pf Suppose $\mathbf{a} \in C\cap C^{\perp}$. Then, by $\mathbf{a}\in C$, there exist $x_{1},\ldots,x_{k}\in R$ such that $\mathbf{a}=\sum_{j=1}^{n}x_{j}G(j,:)$, where $G(j,:)$ is  $j$-th row of $G$.

Since $\mathbf{a}\in C^{\perp}$, we have $$[\sum_{j=1}^{n}x_{j}G(j,:),G(i,:)]=0\mathrm{~for~any}~i\in \{1,2,\ldots,k\},$$  that is $$GG^{T}\mathbf{x}=0,$$
where $\mathbf{x}=(x_1,\ldots,x_k)^{T}$.
It follows that $\mathbf{x}=0$ since $GG^{T}$ is invertible. Hence $C\cap C^{\perp}=\{0\}$, i.e., $C$ is a $\mathbb{Z}_2\mathbb{Z}_2[u]$-LCD code.
\qed

Based on this Theorem we get the  following two corollaries.

\begin{cor} \label{cor:1} Let  $C$ be a  $\mathbb{Z}_2\mathbb{Z}_2[u]$-linear code with generator matrix  $G=\begin{pmatrix}I_{k}&A&0&uB\end{pmatrix}$. If $-1\notin\mathrm{Spec}(AA^{T})$, where $\mathrm{Spec}(M)$ denotes the of all eigenvalues of the matrix $M$, then $C$ is  a $\mathbb{Z}_2\mathbb{Z}_2[u]$-LCD code.
\end{cor}
\pf Since  $-1\notin\mathrm{Spec}(AA^{T})$, we have
$$|GG^T|=|I_k+AA^T|=|(-1)((-1)I_k-AA^T)|=(-1)^k|(-1)I_k-AA^T)|\neq0.$$
Hence, $C$ is  a $\mathbb{Z}_2\mathbb{Z}_2[u]$-LCD code.
\qed

\begin{cor} \label{cor:2}Suppose that $C_1$ is an $[n_1,k,d_H(C_1)]$ binary LCD  code with a generator matrix $G_1$ and  $C_2$  a  $\mathbb{Z}_2\mathbb{Z}_2[u]$ self-orthogonal code of type $(\alpha,\beta,k_0,k_1,k_2)$ with  the standard form matrix $$G_2=\left(
\begin{array}{ccccc}
I_{k_0}& A_{1}&0& 0 &uP\\
0 &S &I_{k_1} &A &B_1+uB_2\\
0 &0 &0 &uI_{k_2}& uD\\
\end{array}
\right),$$  where $k=k_0+k_1+k_2$.  Then the code $C$ with a generator matrix $G=(G_1\mid G_2)$ is a $\mathbb{Z}_2\mathbb{Z}_2[u]$-LCD  code, and   $d_L(C)\geq d_H(C_1)+d_L(C_2)$.
\end{cor}
\pf Clearly, we have  $GG^T = G_1G_1^T$. Since $C_1$ is a binary LCD  code, then $G_1G_1^T$ is invertible by [15, Proposition 1]. And so  $GG^T $ is invertible.
According to Theorem \ref{th:A}, $C$ is a $\mathbb{Z}_2\mathbb{Z}_2[u]$-LCD code. The minimum Lee distance of $C$ follows from the minimum  Hamming  distance of $C_1$ and the minimum  Lee distance of $C_2$.
\qed

\begin{cor} \label{cor:3} Suppose that $C_i$ is a  $\mathbb{Z}_2\mathbb{Z}_2[u]$-LCD code with generator matrix $G_i$  for $i=1,2$. Then the  code $C$ with generator matrix $G=G_1\otimes G_2$ is also a $\mathbb{Z}_2\mathbb{Z}_2[u]$-LCD  code, where  $G_1\otimes G_2$  denotes the Kronecker product of $G_1$ and $G_2$. Moreover,  $d_L(C)= d_L(C_1)d_L(C_2)$.
\end{cor}
\pf Note that $(G_1\otimes G_2)(G_1\otimes G_2)^T = (G_1G_1^T)\otimes(G_2G_2^T)$. Since $G_iG_i^T$ (for $i = 1, 2$) is invertible, we have  $((G_1G_1^T)\otimes(G_2G_2^T))((G_1G_1^T)^{-1}\otimes(G_2G_2^T)^{-1})=((G_1G_1^T)(G_1G_1^T)^{-1})\otimes(G_2G_2^T)(G_2G_2^T)^{-1})=I_{k_1}\otimes I_{k_2}=I_{k_1k_2}$, where $I_a$ is the identity matrix of order $a$.  This means that $(G_1G_1^T)\otimes(G_2G_2^T)$ is invertible. By Theorem \ref{th:A}, $C_1 \otimes C_2$ is a $\mathbb{Z}_2\mathbb{Z}_2[u]$-LCD code.
\qed

The following example shows that   the converse of Theorem \ref{th:A}, in general, does not hold.

\begin{exam}
Let $C$ be a  $\mathbb{Z}_2\mathbb{Z}_2[u]$-linear code of type $(2,3,2,0,0)$ with generator matrix in standard form as follows
$$G=\left(
\begin{array}{ccccc}
1&0&0&u&u\\
0&1&1&u&u
\end{array}
\right).$$

Obviously, the determinant of $GG^{T}$ is equal to zero, i.e., $ GG^{T}$ is not invertible.
We now prove that $C$ is a $\mathbb{Z}_2\mathbb{Z}_2[u]$-LCD code.
In fact, for any $\mathbf{v}\in C$, there are $a_{1},a_{2}\in \mathbb{Z}_2\mathbb{Z}_2[u]$ such that
\begin{equation*}
\begin{split}
\mathbf{v}&=a_{1}(1,0,0,u,u)+a_{2}(0,1,1,u,u)=(\theta(a_1),\theta(a_2),a_2,a_1u,(a_1+a_2)u).
\end{split}
\end{equation*}

Let $\mathbf{v}_{1}=(1,0,0,u,u),~\mathbf{v}_{2}=(0,1,1,u,u)$.
Assume that $\mathbf{w}\in C\cap C^{\perp}$. Then by $\mathbf{w}\in C$ we can find elements $b_1$ and $b_2$ of $R$
such that $\mathbf{w}=(\theta(b_1),\theta(b_2),b_2,b_1u,(b_1+b_2)u).$
Since $\mathbf{w}\in C^{\perp}$, we have $[\mathbf{w},\mathbf{w}]=0$ and $[\mathbf{w}, \mathbf{v}_i]=0$ for  $1\leq i\leq2$.
Hence we have
$$\left\{ \begin{aligned}
        b_{2}^{2}=0, \\
        \theta(b_{1})^{2}+\theta(b_{2})^2=0,\\
        \theta(b_{1})=0,\\
        u\theta(b_{2})+b_{2}=0.
         \end{aligned}\right.$$
The implies  that $b_{2}=0$, $b_{1}=0$ or $u$. Thus $\mathbf{w}=0$, i.e., $C\cap C^{\perp}=\{0\}$. We have shown that $C$ is a $\mathbb{Z}_2\mathbb{Z}_2[u]$-LCD code.
\end{exam}

Under suitable conditions, a converse to Theorem \ref{th:A} holds.
\begin{thm}\label{th: B}
Let  $C\neq\{\mathbf{0}\}$ be a  $\mathbb{Z}_2\mathbb{Z}_2[u]$-linear code with generator matrix  $G$. For $1\leq i\leq k$, if every row $G(i,:)=(g_{i1},\ldots,g_{i,\alpha},\ldots,g_{i,\alpha+\beta})$ there exists  $\alpha<j_i\leq \alpha+\beta$ such that $g_{i,j_i}$ is a unit of $R$. Then $C$ is  $\mathbb{Z}_2\mathbb{Z}_2[u]$-LCD if and only if the $k\times k$ matrix $GG^{T}$ is invertible, where $k$ is the number of rows of   $G$.
\end{thm}
\pf The sufficient part follows from Theorem \ref{th:A}.

The following we prove the necessary condition. Suppose that $GG^{T}$ is not invertible. Then there is a nonzero vector $\mathbf{b}=(\delta_1,\ldots,\delta_k)\in R^{k}$ such that $ GG^{T}\mathbf{b}^{T}=0$. Set $\mathbf{a}=\mathbf{b} G$. If $\mathbf{a}$ is a zero vector of $C$,  then $$\delta_1G(1,:)+\cdots+\delta_kG(k,:)=0.$$
Since $G(1,:),\cdots,G(k,:)$ are  $R$-independent, it follows that $\delta_jG(j,:)=0$ and $\delta_j\in\mathfrak{m}$ for all $j$. This is a contradiction to the assumption. Thus $\mathbf{a}\neq0$.

By  $ GG^{T}\mathbf{b}^{T}=0$, we have also $G\mathbf{a}^T=0$, which is implies that $\mathbf{a}\in C^{\perp}$.

This gives  $C\cap C^{\perp}\neq\{0\}$, i.e.,  $C$ is not a $\mathbb{Z}_2\mathbb{Z}_2[u]$-LCD code.
\qed

Let $C$ is  a $\mathbb{Z}_2\mathbb{Z}_2[u]$-linear code. Define $$C_{\alpha}=\{\mathbf{a}\in\mathbb{Z}_2^{\alpha}|\mathrm{there~exist}~\mathbf{ b}\in
R^{\beta}~\mathrm{such ~that}~(\mathbf{a},\mathbf{ b})\in C\},$$
and
$$C_{\beta}=\{\mathbf{b}\in R^{\beta}|\mathrm{there~exist}~ \mathbf{a}\in \mathbb{Z}_2^{\alpha}
~\mathrm{such ~that}~(\mathbf{a}, \mathbf{ b})\in C\}.$$

It is easy to prove that $G_{\alpha}$ is the generator matrix of $C_{\alpha}$ and $G_{\beta}$ is the
generator matrix of $C_{\beta}$, then  $G=(G_{\alpha}|G_{\beta})$  is  the generator matrix of the $\mathbb{Z}_2\mathbb{Z}_2[u]$-linear code $C$.


Let $C$ be  a $\mathbb{Z}_2\mathbb{Z}_2[u]$-linear code. If $C = C_{\alpha}\times C _{\beta}$ , then $C$ is called separable.

Let $C_{\beta}$ be a linear code of length $\beta$ over $R$. If $C_{\beta}\cap C_{\beta}^{\perp}=\mathbf{0}$, then $C_{\beta} $ is said to be an $R$-LCD code.

\begin{thm} \label{th:3.10} Let $C$ is  a $\mathbb{Z}_2\mathbb{Z}_2[u]$-linear code. If $C_{\alpha}$ and $C_{\beta}$ are binary LCD and $R$-LCD codes,respectively, then  $C$ is a $\mathbb{Z}_2\mathbb{Z}_2[u]$-LCD code.
\end{thm}
\pf Suppose that $C$ is  not a $\mathbb{Z}_2\mathbb{Z}_2[u]$-LCD code. Then exist $0\neq\mathbf{v}=(\mathbf{x},\mathbf{y})\in C\cap C^{\perp}$. We divide the rest of the proof into two cases.

Case 1. If $\mathbf{x}\neq0$, then by $\mathbf{v}=(\mathbf{x},\mathbf{y})\in C\cap C^{\perp}$ we have $\mathbf{x}\in C_{\alpha}\cap C_{\alpha}^{\perp}$, which is a contradiction.

Case 2. If $\mathbf{y}\neq0$, then by $\mathbf{v}=(\mathbf{x},\mathbf{y})\in C\cap C^{\perp}$ we have $\mathbf{y}\in C_{\beta}\cap C_{\beta}^{\perp}$, which is also a contradiction.
\qed

\begin{cor} Let $C = C_{\alpha}\times C _{\beta}$. Then $C$ is a $\mathbb{Z}_2\mathbb{Z}_2[u]$-LCD code if and only if $C_{\alpha}$ and $C_{\beta}$ are binary LCD and $R$-LCD codes,respectively.
\end{cor}

The reverse statements of Theorem \ref{th:3.10} is not true in general.

\begin{exam}
Let $C$ be a  $\mathbb{Z}_2\mathbb{Z}_2[u]$-linear code 
generated by
$$G=\left(
\begin{array}{cccccc}
1&0&0&1&0&0\\
0&1&0&0&1&0\\
0&0&1&0&0&1\\
1&1&1&u&u&u\\
\end{array}
\right).$$

Obviously, $C_{\alpha}=\mathbb{Z}_2^{\alpha}$ and $C_{\beta}=R^{\beta}$ are binary LCD and $R$-LCD codes, respectively.
However, the last row $\mathbf{a} = (1~1~1~u~u~u )$ is orthogonal to any row in the generator matrix.
Hence, $\mathbf{a}\in C \cap C^{\models}$ and C is not a $\mathbb{Z}_2\mathbb{Z}_2[u]$-LCD code.
\end{exam}

\begin{thm} \label {th:F} A $\mathbb{Z}_2\mathbb{Z}_2[u]$-linear code $C$ is $\mathbb{Z}_2\mathbb{Z}_2[u]$-LCD if and only if $\Phi(C)$ is a binary LCD code.
\end{thm}
\pf We firstly prove that $\Phi(C\cap C^{\perp})=\Phi(C)\cap \Phi(C^{\perp}).$

In fact, if $\Phi(\mathbf{w})\in\Phi(C\cap C^{\perp})$, then $\mathbf{w}\in C\cap C^{\perp}$. Thus,  $\Phi(\mathbf{w})\in\Phi(C)$, and $\Phi(\mathbf{w})\in\Phi(C^{\perp})$, which is implies that $\Phi(\mathbf{w})\in\Phi(C)\cap \Phi(C^{\perp})$. Therefore,
$$\Phi(C\cap C^{\perp})\subset\Phi(C)\cap \Phi(C^{\perp}).~~~~~~~~~~~~~~~~(3.1)$$

On the other hand, suppose that $\mathbf{v} \in \Phi(C)\cap \Phi(C^{\perp})$, then there exist $\mathbf{x}\in C$ and $\mathbf{y}\in C^{\perp}$ such that
$\mathbf{v}=\Phi(\mathbf{x})=\Phi(\mathbf{y})$. Since $\Phi$ is an isomorphism, we have $\mathbf{x}=\mathbf{y}\in C\cap C^{\perp}$. Thus, $\mathbf{v}\in\Phi(C\cap C^{\perp})$.  This means that
$$\Phi(C\cap C^{\perp})\supset\Phi(C)\cap \Phi(C^{\perp}).~~~~~~~~~~~~~~~~~~~~~~~~~(3.2)$$
Combining $(3.1)$ and $(3.2)$, we get the desired equality.

Now, we assume that a $\mathbb{Z}_2\mathbb{Z}_2[u]$-linear code $C$ is $\mathbb{Z}_2\mathbb{Z}_2[u]$-LCD, then by  $\Phi(C\cap C^{\perp})=\Phi(C)\cap \Phi(C^{\perp})$ and $\Phi(C^{\perp})=\Phi(C)^{\perp}$,  we  know that $\Phi(C)$ is a  binary LCD code over $\mathbb{Z}_2$.

Conversely, let $\Phi(C)$ be is a binary LCD code.  Assume that $C$ is not a $\mathbb{Z}_2\mathbb{Z}_2[u]$-LCD code, then there exists $0\neq \mathbf{x}\in C\cap C^{\perp}$ such that
$$\Phi(\mathbf{x})=(a_0,\ldots,a_{s-1},c_0,\ldots,c_{t-1},b_0+c_0,\ldots,b_{t-1}+c_{t-1})\in \Phi(C)\cap \Phi(C^{\perp}).$$
where  $\mathbf{x}=(a_0,\ldots,a_{s-1},b_0+uc_0,\ldots,b_{t-1}+uc_{t-1})$.

We assert that $\Phi(\mathbf{x})\neq0$. Otherwise, by $\Phi(\mathbf{x})=0$, we have
$$a_0=\cdots=a_{n-1}=0,c_0=\cdots=c_{n-1}=0,$$
and $$b_0+c_0=0,\ldots,b_{t-1}+c_{t-1}=0.$$
This means that $\mathbf{x}=0$. This is a contradiction.

Therefore, $0\neq\Phi(\mathbf{x})\in \Phi(C)\cap \Phi(C^{\perp})$, which is a contradiction to the  assumption.
\qed

\section{Examples of new binary LCD codes}
In this section, examples of  some new binary codes derived from this family of $\mathbb{Z}_2\mathbb{Z}_2[u]$-LCD codes as their Gray images are presented.

\begin{exam}Let $C$ be a $\mathbb{Z}_2\mathbb{Z}_2[u]$-linear code of type $(9,9)$ with the generator matrix
$$G=\left(
\begin{array}{cccccccccccccccccc}
1&1& 0& 1& 0 &1& 1 &0&1& 1 + u &1 + u &1 + u& 1 &1& 1& 1 &1& 1\\
1& 0 &1& 1 &1& 1& 0& 1 &0& u &0& 0& u& 0 &0& 0& 0 &0\\
0 &1& 0& 1 &1& 1& 1& 0& 1& 0& u& 0& 0 &u& 0& 0& 0 &0\\
1& 0& 1& 0& 1 &1& 1& 1& 0& 0& 0& u& 0& 0 &u &0 &0 &0\\
0& 1& 0& 1 &0& 1& 1 &1 &1& 0& 0& 0& u& 0&0 &u& 0& 0\\
1& 0& 1& 0 &1& 0& 1& 1& 1& 0& 0 &0& 0 &u &0& 0&u&0\\
1& 1& 0 &1 &0& 1& 0& 1& 1& 0 &0 &0& 0& 0& u& 0& 0& u
\end{array}
\right).$$
Obviously,
$$GG^T=\left(
\begin{array}{ccccccc}
1 1 1 1 1 1 1\\
1 0 1 1 1 0 0\\
1 1 0 1 1 1 0\\
1 1 1 0 1 1 1\\
1 1 1 1 0 1 1\\
1 0 1 1 1 0 1\\
1 0 0 1 1 1 0\\
\end{array}
\right).$$
Thus, $det(GG^T)=1$.  By Theorem \ref{th:A}, the $\mathbb{Z}_2\mathbb{Z}_2[u]$-linear code $C$ is a $\mathbb{Z}_2\mathbb{Z}_2[u]$-LCD code.  Again by Theorem \ref{th:F}, the binary image of this code is a binary LCD code with parameters $[27,8,10]_2$. This is an optimal code which is obtained directly in contrast to the indirect constructions presented in \cite{Gra1}.
\end{exam}

\begin{exam}

Let $C$ be a $\mathbb{Z}_2\mathbb{Z}_2[u]$-linear code of type $(7,7)$ with the generator matrix
$$G=\left(
\begin{array}{cccccccccccccccccc}
1&1& 0& 1& 0 &0& 0 &0&0& 0 &0 &0& 0&0\\
0&1&1& 0& 1& 0 &0& 0 &0&0& 0 &0 &0& 0\\
0&0&1&1& 0& 1& 0 &0& 0 &0&0& 0 &0 &0\\
0&0&0&1&1& 0& 1& 0 &0& 0 &0&0& 0 &0 \\
1& 1&  1 &0& 0& 0 &0 &1+u&1+u& 1+u& 1& 1& 1+u&1\\
1& 0& 0& 1 &0& 0&0& u& u& u& 0 &u& 0 &0\\
0&1& 0& 0& 1 &0& 0&0& u& u& u& 0 &u& 0 \\
0&0&1& 0& 0& 1 &0& 0&0& u& u& u& 0 &u
\end{array}
\right).$$
Obviously, $det(GG^T)=1$.  By Theorem \ref{th:A}, the $\mathbb{Z}_2\mathbb{Z}_2[u]$-linear code $C$ is an $\mathbb{Z}_2\mathbb{Z}_2[u]$-LCD code.  Again by Theorem \ref{th:F}, the binary image of this code is a binary LCD code with parameters $[21,8,3]_2$.
\end{exam}

\begin{exam}
Let $C$ be a $\mathbb{Z}_2\mathbb{Z}_2[u]$-linear code of type $(31,31)$ with the generator matrix
$$G=\left(
\begin{array}{ccccccccccccccccccccccccccccccccccccccccccccccccccccccccccccccccccccc}
1011010100011101111100100010000u0uu0u0u000uuu0uuuuu00u000u0000\\
01011010100011101111100100010000u0uu0u0u000uuu0uuuuu00u000u000\\
001011010100011101111100100010000u0uu0u0u000uuu0uuuuu00u000u00\\
0001011010100011101111100100010000u0uu0u0u000uuu0uuuuu00u000u0\\
00001011010100011101111100100010000u0uu0u0u000uuu0uuuuu00u000u
\end{array}
\right).$$

Obviously,
$$GG^T=\left(
\begin{array}{ccccc}
1&1 &0& 1 &0\\
1 & 1& 1  &0&1\\
0& 1  &1 &1 & 0\\
1 &0& 1& 1& 1\\
0 &1  &0 &1& 1
\end{array}
\right).$$
Thus, $det(GG^T)=1$.  By Theorem \ref{th:A}, the $\mathbb{Z}_2\mathbb{Z}_2[u]$-linear code $C$ is an $\mathbb{Z}_2\mathbb{Z}_2[u]$-LCD code.  Again by Theorem \ref{th:F}, the binary image of this code is an optimal binary LCD code with parameters $[93,7,46]_2$.
\end{exam}

\begin{exam}
Combining  Corollary \ref{cor:2} and Theorem \ref{th:F}, we obtain some new binary LCD codes in  Table 1.
\begin{table}
\caption{New  binary LCD codes obtained from Corollary \ref{cor:2}}
\begin{center}\begin{tabular}{|c|c|c|}
\hline
 self-orthogonal codes in Ref \cite{Ay} &  LCD codes in Ref\cite{Hara},\cite{Sok},\cite{Gali},\cite{Rao} & New  binary LCD codes\\
\hline
$[18,2,8]_2$& $[10,2,6]_2$,&$[28,6,\geq 14]_2$\\& $[12,2,7]_2$,&$[30,2,\geq 15]]_2$,\\&$[13,2,8]_2$,&$[31,2,\geq 16]_2$,\\&$[14,2,9]_2$ &$[32,2,\geq 17]]_2$,\\&$[15,2,10]_2$,&$[33,2,\geq 18]_2$\\\hline
$[56,25,6]_2$ & $[35,25,4]_2$ & $[91,25,\geq 10]_2$\\\hline
$[56,28,6]_2$ & $[35,28,\geq3]_2$& $[91,28,\geq 9]_2$\\& $[65,28,14]_2$,&$[121,28,\geq 20]_2$\\&$[66,28,14]_2$,&$[122,28,\geq 20]]_2$\\\hline
$[98,48,6]_2$ & $[56,48,\geq3]_2$,& $[154,48,\geq 9]_2$\\& $[63,48,6]_2$,&$[161,48,\geq 12]_2$\\&$[257,48,74]_2$,&$[355,48,\geq 80]_2$\\\hline
\end{tabular}\end{center}
\end{table}
\end{exam}

\section{Conclusion}
We have developed new methods of constructing  binary LCD codes from  $\mathbb{Z}_2\mathbb{Z}_2[u]$-codes. Using those methods, we have constructed  good  binary LCD codes. We believe that  $\mathbb{Z}_2\mathbb{Z}_2[u]$-LCD codes will be a good source for constructing good binary LCD codes. In a future work, in order to construct new  binary LCD,   we will use the computer algebra system MAGMA to find more good $\mathbb{Z}_2\mathbb{Z}_2[u]$-LCD codes.

\end{document}